# LANGUAGE, PARTICIPATION AND INCLUSIVITY IN URBAN PLANNING PROCESSES IN MZUZU CITY, MALAWI

**Francis Engwayo Mgawadere[1] and Mtafu Manda[2]**

[1]*Dept of Humanities, University of Livingstonia, P.O. Box 37, Laws Campus, Livingstonia, Malawi. Email: fmgawadere@unilia.ac.mw* Cell: +265 992014094

[2] *Department of the Built Environment, Mzuzu University, Private Bag 201, Mzuzu, Malawi. Email: manda.ma@mzuni.ac.mw*  Cell: +265 991457275

## ABSTRACT

*Participation in urban planning is championed for entrenching democracy and development. Malawi passed the Local Government Act (1998) and Decentralization Policy (1998) to facilitate community participation in decision-making processes. Several studies have been conducted on decentralization and local governance on community participation. Little attention has been paid to examining the impact of the language used in planning processes on democracy and inclusivity envisaged in the law and policy. Using communicative action theory, the study examined challenges posed by language used in planning processes on inclusivity in the approval processes of urban plans. Data were collected through semi-structured interviews, focus group discussions, observations and document review and analyzed using thematic and discourse analysis. The findings show that while there is high participation at community planning levels, because planners communicate using local languages, participation is compromised in the service committees at city level where final planning decisions are made due to language barrier. Specifically, lack of sincerity, truthfulness, comprehensibility and therefore legitimacy are apparent. Planners are reluctant to simplify written language and translate planning jargon into local languages for councillors to understand. The study concludes that community participation in the urban planning process in Mzuzu fails to entrench democracy due to lack of inclusiveness owing to the language barrier at city level where final planning decisions are made. The study proposes a framework for inclusive participation in urban planning including the motivation, conditions for effective participation and outcomes of participation.*

Community participation in decision-making in the urban planning process has been championed for entrenching democratic ideals and development outcomes. Participation was defined by the United Nations Research Institute for Social Development (UNRISD) as: *"the organized efforts to increase control over resources and regulative institutions in given situations, on the part of groups and movements hitherto excluded from such control"* (Stiefel & Wolfe, 1994: 5). The Malawi Government passed the Local Government Act (1998 amended 2024) and the National Decentralization Policy (1998 amended 2024) to facilitate community participation in decision-making in planning and development. Several studies have been conducted on decentralization and local governance in relation to community participation. Little attention has been paid to examining the impact of the language used in planning processes on inclusivity to realize the democratic ideals and development outcomes envisaged in the law and policy. Using the Habermasian communicative action theory, the study examined the challenges posed by the language used in the planning process on inclusivity in the approval processes of urban development plans. Specifically, the study evaluated the influence of planning language on participatory democracy and inclusiveness, by examining the extent to which Habermas precondition of communication, also known as validity claims, namely: a) comprehensibility, b) sincerity, c) truthfulness and d) legitimacy have been met adequately. The failure to meet Habermas validity claims implies failure to entrench democracy and inclusiveness, because communication is itself a precondition of democracy (Taylor, 1998). The paper is structured as follows: section 2 presents literature review. Section 3 presents methodology. Section 4 presents results and discussion. Section 5 presents conclusion and the proposed framework for inclusive participation in urban planning.

Habermas developed a general theory that provides a platform to critique the contemporary capitalist society, while providing the preconditions for a more democratic society, which later inspired the Habermasian theory of Communicative Action (Taylor, 1998). According to Habermas, if two or more people are to communicate effectively with each other, certain conditions have to be met, which he termed; "general presuppositions of communication" (Habermas, 1979: 1). Habermas suggests that, when person A communicates with person B, A implicitly assumes or makes four validity claims; first, A assumes that what he is saying is, comprehensible (i.e., understandable) to B. This is obviously a precondition of communication because, if what A is saying is incomprehensible to B, then clearly no communication is taking place between A and B (Habermas, 1979; 1987; Taylor, 1998). Secondly, for A to communicate to B, it must be A, himself who communicates, from which Habermas infers that A must be "Sincere" in communicating to B. According to Habermas (1987; 1979), the validity claim of sincerity is that for genuine communication to occur between two persons, the speaker must not deceive the listener. Thirdly, A must communicate 'something' to B, from which Habermas infers that A must assume or make the validity claim that the 'something' he is saying is factually "True" (Taylor, 1998: 123). Fourthly, in order for A to communicate to B, A must be seeking to come to an understanding with B. Thus, A must assume that what he is saying is legitimate, within the context of moral norms and conventions shared by both A and B (Habermas, 1979: 2-3). Habermas (1979) argues that the communicative action theory enables us to envisage the four preconditions of communication as: *comprehensibility, truthfulness, sincerity, and legitimacy* (Taylor, 1998). If these four preconditions cannot be met, then no genuine communication will take place. As communication is itself a preconditions for real democracy, and hence, of any democratic participation in planning, and without genuine communication, there can be no genuine participation in urban planning and decision-making (Taylor, 1998).

The leading pioneer of the communicative planning theory has been an American, known as John Forester, who has drawn extensively on Habermasian theory as a vehicle for evaluating planning practice in terms of the ideals of good communication and democratic participation (Taylor, 1998). In his 1989 book; *Planning in the Face of Power*, Forester begins from the premises that "planning is for the people" and in Western liberal democracies, the planning practice is constrained by the political realities of a capitalist society (p. 3). His aim is to explore the skills that planners need to maximize their effectiveness in planning for people in the face of power (Forester, 1989). He asserts that in order to get things done, planners have to be effective communicators and negotiators, because in planning, talk and argument matter and that the daily business of a planner is basically communicative (Forester, 1989: 5 & 11). He insists that in getting things done, urban planning should aspire to the ideals of democratic decision-making over the development proposals (Forester, 1989). While planners will be negotiating with powerful developers, they should also be active in protecting the interests of all groups in the society, including the less powerful or marginalized communities (Forester, 1989). Drawing from Habermas, Forester emphasizes the duty of planners to facilitate participatory democracy in planning. By emphasizing planner's duty to involve the less powerful groups, by exposing distorted communication and misinformation, Forester sees planning as a communicative process carrying with it, a communicative ethos (Forester, 1989: 22-24).

Further, the fourth validity claim number 3: the plans should be righteous – meaning that the plans should be right when in accordance to planning standards, life expectancy and access to resources are maintained; and opens new possibilities in politics and social sciences". The origin of the prevailing communicative planning theory is the pragmatic approach by Habermas.

However, in his study; '*The Dark Side of Planning: Rationality and Real Rationalitat*', in

According to Dados & Cornell (2012; 13), "It references an entire history of colonialism, neo-standards, life expectancy and access to resources are maintained; and opens new possibilities in politics and social sciences". The origin of the prevailing communicative planning theory is the pragmatic approach by Habermas.

A qualitative research approach was employed for this study in order to investigate social interactions and explore meanings that communities ascribed to their participation in the urban planning process. As the philosophical stance of this study is interpretive, the qualitative methods enabled the researcher to explore the multiple constructions of reality from the diverse opinions in order to understand the phenomenon (Bryman, 2004; Myers, 2008).

Mark

The study used the multi-stage cluster sampling, to divide the study population into smaller groups starting from the members at block level, to neighbourhood and ward committees and all the way up to Council Service committees at Mzuzu City Council. Purposive sampling was used to select members of these groups for an in-depth and key informant interviews (KIIs), focus group discussions (FGDs) and observation.

Data was collected through semi-structured interviews, FGDs, observations of planning engagements and document review. Qualitative data from community members, such as, block leaders, members of the neighbourhood and ward committees, were extracted through in-depth interviews and focus group discussions (FGDs). Qualitative data from key informants such as councillors, planners and planning officials at Mzuzu City Council, were collected by administering key informant interviews (KIIs) and observation. Qualitative secondary data was also collected from key planning documents such as urban development plan (UDP), urban structure plan (USP) and planning and service committee minutes.

Data was analyzed using deductive thematic analysis and discourse analysis. Unlike inductive thematic analysis (a qualitative research methods where themes emerge directly from the data), the researcher used pre-defined themes derived from Habermas validity claims of ***comprehensibility, sincerity, truthfulness and legitimacy***, to confirm or refute them. The researcher also used discourse analysis to analyse planning language within its social context, in terms of how planners speak to communities. By using the discourse analysis, the researcher was able to identify how power dynamics between communities influence meanings of the planning concepts.

The data enabled the researcher to explore the extent to which planning language influences the entrenchment of democracy and inclusiveness in the decision-making processes in urban planning in Mzuzu City. The study focused on confirming or refuting the pre-defined themes drawn from Habermas validity claims of comprehensibility, sincerity, truthfulness and legitimacy.

This theme investigated the extent to which planning language is understandable to communities when planners communicate advice, information and knowledge orally and in writing. The findings show that the Habermas validity claim of comprehensibility has not been met. There is a higher level of understanding between planners and communities at the community planning level than in the Council Service Committees at City level, because at the community level, planners speak the local language and all communities understand the communication:

> "*When the conversation is done in local languages, many people speak a lot of sense…*" **(KII-CLLR1/18-11-24).** "*We understand them 100% because they speak to us in local languages*" **(KII-CLLR2/14-11-24)**.

These quotes reveal that community participation is higher at the block level, neighbourhood and Ward development committee levels. At the community planning level, communities are required to identify, select and prioritize development projects that they need. It is the first level of planning. Final planning and funding decisions not are made here.

However, participants said that community representatives (councillors) struggle to follow the deliberations in the council service committees at City level, due to language barrier:

> *…the language is English …councilors do not understand the language… using jargons which are not of their area. Attempts are there to orient them on the operations, they know what to put, but you find that by the end of the day the language still needs to be simplified for them* **(KII-P1/14-10-24).**

> *Not all of us* [councillors] *understand English. Most of the information about planning and development is spoken in English. So, most people do not understand… The deliberations in all service committees are conducted in English* **(KII-CLLR1/18-11-24).**

The extracts reveal that the level of understanding is low because councillors who represent communities in the Council Service Committees fail to follow the deliberations because many of them do not fully comprehend English. The problem gets worse when technocrats use technical language, planning and financial jargons. The study revealed evidence of reluctance to simplify spoken language for community representatives' ease of understanding:

> *"And that's why, I have a fight with the Secretariat, because… [They] don't want a learned person ….. They want these councilors with lower education so that when they present the reports, councilors do not understand…."* **(KII-P4/22-11-24).**

These findings do not agree with Habermas (1979; 1987) who stresses the need for planning language to be understandable to hearers. The failure of councillors to understand and follow the deliberations in the Physical Planning, Finance and other committees implies that grassroots communities are not participating and certainly not being included in the decision-making processes that lead to final decisions. Evidence from the minutes of the Physical Planning and Finance Committees, also indicate that councillors are just passive listeners.

Further, the study also found that key planning documents like the urban development plans (UDP) and the urban structure plans (USP) have not been translated into the local languages for the majority to understand. There was a perceived reluctance to translate technical language, planning jargons and documents into the local language:

> *"...during community outreach programmes, we present ourselves in local languages… but to date the Five Year-Urban Development Plan has not yet been translated… [Of course] translating the whole document would not be meaningful for me because not every information… is relevant to a local person"* **(KII-P3/29-10-24).**

However, communicative planning requires planners to use an easy-to-understand language in planning texts, to simplify planning and technical jargons and even translate them into vernacular languages. For instance, Forester, (1989: 149) advises planners to present their ideas and information in a manner that is less obscure, comprehensible or easy to understand. Habermas (1979; 1987) emphasizes the need for language to always meet the validity claim of comprehensibility, meaning that information being presented by speakers (planners) should be understandable to hearers (communities). The failure of communities to understand what planners are communicating implies that communities are manipulated. It is in the Council Service Committee meetings that crucial planning and funding decisions are made before being forwarded to the Full Council for approval. The fact that inputs from communities are lacking means that the policies and city by-laws do not reflect the will of the majority or the common people. Thus, planning is not democratic and inclusive in Mzuzu City.

The results of this investigation do not align with the Habermas (1979; 1987) validity claim of comprehensibility. According to Habermas (1979), if A communicates to B, he must assume that what A is saying is comprehensible to B. However, in this study, what A (planners) were saying could only be understood by B (communities) at the block and ward level, where communities only identify, select and prioritize development projects, because the local language is used for communication. However, when the prioritized projects are sent to service committees, councillors (B), fail to participate effectively in the decision-making process about planning and funding allocation, due to language barrier, as A, uses English that is full of technical and planning jargons and the planning texts are not translated into the local language for communities to understand easily.

These two overlapping and similar themes investigated the extent to which planners' oral and written communication is honest, less deceptive and truthful in order to determine whether planning language enhances participation and inclusivity in the decision-making process in urban planning.

First, the validity claim of sincerity investigated levels of honesty or deception in planners' communication. The findings indicate that while the language planners use to communicate with communities appears to be superficially honest and less deceptive during planning meetings, but many interviewees complained that planners are deceptive and dishonest during project implementation:

> *They sound sincere …but deceptions arise during budgeting and resource allocation. Selection and identification of contractors is done by themselves. Councillors are not involved… the Internal Procurement Committee (IPC) sits down to discuss bids on their own. There are no community representatives in the IPC… This is where deceit comes in because they select a contractor who promises kickbacks* **(KII-CLLR2/14-11-24).**

> *They speak to us very sincerely during planning meetings. The problems arise during implementation. This is when I think they indulge in fishy businesses. The process gets messed up during the identification and awarding of contracts* **(KII-CLLR1/18-11-24).**

*No. they are not sincere. There is a kind of deception. Mostly about 30% of the language is deceitful. Largely, deceit comes in so that they can easily convince and walk through. They use deception to advance their own ulterior motives* **(KII-P1/14-10-24).**

These extracts reveals how deceitful planners are. There is no transparency and accountability when it comes to selecting project contractors. This is done in the internal procurement committee (IPC) in the absence of community representatives. So, while planners sound very sincere during planning meetings, but after that, they implement different things that are contrary to what they communicated publicly. Further, the study revealed evidence of reluctance on the part of the Secretariat to simplify spoken language to enable councillors in service committees understand when financial reports and statements are being presented. In an interview, the City Mayor revealed that he sometimes fights with the Secretariat for resisting concerns about councillors' failure to follow deliberations in the service committees because their level of education is low. He said that the Secretariat is aware that councillors do not understand presentations of the financial reports and statements and that the use of technical language and planning jargons is a deliberate ploy to conceal crucial information from public scrutiny. Also, the key planning documents such as the Urban Development Plan (UDP) and the Urban Structure Plan (USP) are professionally written and that there is no evidence of insincerity and untruthfulness, but problems arise during implementation. This is when what has been communicated according to Habermas criterion, gets distorted during implementation.

Second, the theme of truthfulness investigated the extent to which planners' oral and written communication is factually accurate or truthful. Participants revealed that planners' language is generally truthful when they speak during planning meetings and in planning texts, but there are instances when planners cheat communities:

*… [planners] always sound very truthful when they speak to us… but the problems …arise during budgeting, resource allocation … [when] members of the Internal Procurement Committee sit down to discuss bids, there is no community representative ….what is implemented is different from what they agreed with people…….”* **(KII-CLLR2/14-11-24)**

*Decentralization is just a myth. The real powers are still at the Council level… Usually if it is coming with resources they withhold information from the grassroots communities because they would want to hide financial resources. They don't want to diverge more information to*

*communities because it will make it so difficult to play fishy businesses. The Council gets more meat and give bones to the communities* **(KII-P4/22-11-24).**

These results are consistent to Healey's (1995: 259) assertions that verbal agreements reached according to Habermas validity claims can still be distorted in writing by planners in their offices. Flyvbjerg (1996: 392) questions the idea of viewing 'planners as noble individuals' due to failure to 'speak truthfully'. As further proof of lack of sincerity and truthfulness, the researcher was not allowed to observe deliberations in the council service committees, including the Physical Planning Committee, the Finance Committee and the Internal Procurement Committee, despite prior approval to the letter requesting for consent and repeated requests to observe these committees.

The lack of both sincerity and truthfulness impedes a genuine flow of communication which prevents communities from participating in the decision-making process and thus compromises the goal of entrenching democracy and inclusive urban planning as envisaged in the Local Government Act (1998 amended 2024) and the Malawi decentralization policy (UNDP, 2000; Malawi Government, 1998). This is also contrary to Habermas (1979; 1987) emphasis that the language should always meet the validity claim of comprehensibility, meaning that information being presented by speakers should be understandable to hearers, the validity claim of sincerity which requires higher levels of honesty and the validity claim of truthfulness which demand higher levels of factual accuracy. This apparent lack of honesty and truthfulness impedes a genuine flow of communication between planners and communities, and prevents inclusive community participation in decision-making processes over budgeting, resources allocation, determination of planning applications for development permission and the selection of project contractors in the IPC. This also means that the final budgeting and funding decisions are devoid of the inputs from the grassroots communities thereby, rendering the planning process less participatory and not inclusive.

On the one hand, the findings agree with Flyvbjerg (1996) study entitled; **'The Dark Side of Planning: Rationality and the Real Rationalitat'**, in Aalborg, Den Mark, a European country in the Global North. Flyvbjerg (1996) found the idea of viewing planners as noble creatures as a myth, that planners are unethical, crooks, liars, deceivers and corrupt professionals, that the reality in which planning takes place is false, cruel, contradictory and seductive, and that planners, as human beings need lies to survive (Flyvbjerg, 1996: 391). He insists that most observers would agree that deception is part and parcel of many everyday decisions in government because the social incentives of deception are at present very powerful, while controls are often weak, and that deceptions are part and parcel of the decisions planners are involved in and that the incentives for planners to deceive others are strong (Flyvbjerg, 1996: 392). He concludes that the idea that planners are noble individuals with good manners are plain lies of the planning theorists (Flyvbjerg, 1996: 391).

On the other hand, these results do not align and spits in the face of Henri Lefebvre's (1996) influential ideas about the 'right to the city', in which he states that cities should be understood as common goods, benefiting all residents, rather than just the rich and powerful oligarchs, that cities should be inclusive, where every street, every building and every corner belongs to the people who live there, not just the rich, the planners and the entitled, but all of us. He criticized technocratic planning, arguing that the common people should have power to shape the cities they live in and that cities are not just spaces for power and control, but are living spaces that we create together, yet too often they are controlled by capitalist oligarchs, profit hungry developers and apathetic governments (Lefebvre, 1996).

This theme investigated the extent to which planners, oral and written communication meets Habermas (1979; 1987) validity claim of legitimacy. Legitimacy is concerned with how communication complies with the normative values and conventions (Forester, 1993).

The findings from a review of the planning language in the three key planning texts; urban structure plan (USP), urban development plan (UDP) and urban profile (SEP) for Mzuzu City, indicate that communication complies with the normative values, conventions and laws as outlined in the legal provisions of the Constitution of the Republic of Malawi (1995), the Local Government Act (1998 amended 2024) and the National Decentralization Policy (1998 amended 2024) and the Town and Country Planning Act (1988), (Mzuzu City Council, 2023 – 2030). The language in the three key planning documents also comply with the provisions of the Constitution of the Republic of Malawi of 1995, which requires the full participation of the grassroots communities in the decision-making processes in order to entrench democracy in Malawi. The language also emphasizes the decentralization of powers from Central to local authorities and from the top half to the lower half of the Council:

> *The Grassroots Participation Process (GPP) is the bottom-up process, which involves consultation with the communities that aims to gather information on their needs. A GPP task force is formed whose members work in collaboration with Ward Development Committee; block leaders perform the GPP (NAP process). The output of this is the prioritized list of projects. The resultant output of the urban socio-economic profile and the GPP is the formulation of the Urban Development Planning Framework, which highlights major issues, potentials and development objectives and strategies. The framework forms the basis for the formulation of projects and programmes* **(Mzuzu City (UDP), 2023-2030: p. VIII).**

The excerpt underscores that community participation is central to the urban planning process in Mzuzu City. The urban profile document assessed the current situation and identified available developmental potential for the Council and using these findings, planners developed the urban development strategy, programmes and projects to be implemented in the next 10 to 15 years. The formulation of the UDP and USP was based on the findings of the process of dialogue, inclusive democracy and discourse with equal distribution of power for argumentation with communities. Thus, planning texts are legitimate because they are based on the outcome of the communicative process and in compliance with the legal provisions of Malawi.

A review of the key planning texts reveals that written communication complies with Habermas (1979) validity claim of legitimacy, which requires the language that planners use to communicate planning ideas, knowledge and information, conform to the normative values, moral conventions and legislation. The three planning texts were written professionally and complies with the legal provisions outlined in the Malawi Constitution, the Local Government Act and the national decentralization policy documents outlined above. All the three key planning documents emphasize participation and representation of the grassroots communities as a way of entrenching participatory democracy and inclusive urban planning.

However, in spite of this compliance, participants lamented that problems still arise in the service committees and in the IPC at City level where final decisions are made in absence of community representatives, as alluded to earlier, and during the implementation phase. The study revealed that while planners may communicate orally that a particular area has been designated as a residential area, in conformity with what is indicated in the urban structure plan (USP), but during implementation, the area designated as residential turns out to be a mixed–use zone of residential and commercial structures and a high density area becomes a mixed zone of high, medium and low density housing structures:

> *The structure plan – they approved it, to say, by law this is zoned for residential and the like, but when it comes to implementation, it becomes different – a mixture of residential, commercial and etc... So the documentation itself is legitimate, but here, you are talking of oral – they are communicating orally legitimate things, but when implementing, it's different - it's illegitimate. Thus, legitimacy is really there. They communicate according to the law – but when it comes to implementation, this legitimacy ceases* **(KII-P1/14-10-24).**

This tendency implies that the actual physical infrastructure development is not in line with what is indicated in the Urban Structure Plan (USP) and the Urban Development Plan (UDP), resulting into a kind of development that reflects the ulterior motives of the rich and powerful oligarchs, instead of the will of the majority of the Mzuzu City residents. This also means that the development of the City is illegitimate because it fails to comply with the legitimate urban plans which were formulated in compliance with the legal provision that emphasize participation and representation of communities in planning and decision-making.

These findings are consistent with Healey (1993) who reveals that verbal agreements reached according to Habermas (1987; 1979) validity claims can still be distorted in writing by planners in their offices. Thus in theory, planners communicate in accordance with the provisions of the laws, which require the participation and representation of communities in all decision-making processes and the decentralization of powers to grassroots communities (Malawi Government, 1995; 1998; UNDP, 2000). However, in in practice, all decisions are made by technocrats in service committees at City level and planners resist delegating these powers to community planning committees despite the decentralization policy requiring them to do so. The fact that councillors who represent their electorate are passive listeners in Council service committees, implies that their communities are not actively participating and influencing planning decisions. They do not actively participate in the deliberations that lead to final planning decisions. They are thus placated and manipulated.

On the one hand, these findings are consistent with those of Yiftachel (1998). In his article, **'Planning and Social Control: Exploring the Dark Side'**. Yiftachel (1998) argues that urban planning, despite its potential for positive change, has a hidden dark side, where it functions as a tool for social control and oppression, especially for the marginalized groups. Influenced by Michel Foucault, Yiftachel (1998) highlights how planners can reinforce existing power structures by manipulating space and socioeconomic conditions to benefit certain groups while excluding others. In the same vein, the results reveal a hidden dark side, where planners exclude community representatives from the decision-making processes about planning, budgeting, resource allocation and the selection of the project contractors.

On the other hand, the findings are contrary to Habermasian theories of communicative rationality and action. Habermas asserts that if two or more people are to communicate effectively, certain conditions have to be met (Habermas 1979; 1). One of these preconditions is legitimacy. According to Habermas, if A communicates to B, he should assume that what A is saying is legitimate (complies with moral norms, conventions and laws) (Habermas 1979; 1; Taylor 1998; 123). The failure to meet the validity claim of legitimacy implies that no genuine communication is taking place, and thus a lack of community participation in urban planning. As genuine communication is a pre-condition for participatory democracy, its absence means the absence of participatory democracy and inclusivity in urban planning.

The limitation of the findings is that the researcher was not granted access to observe deliberations in service committee meetings. However, the researcher concludes that this was further proof of lack of sincerity and truthfulness, because he was not granted access to observe Council service committee deliberations, despite consenting to this earlier on (see the Request for Consent Letter on the Appendix Section). Nevertheless, the researcher managed to access the minutes of the previous service committee meetings and was able to draw conclusions.

### *onclusion*

The study concludes that planning language in the decision-making process in urban planning in Mzuzu City fails to enhance community participation and inclusivity, due to lack of inclusiveness owing to the language barrier at city level where final planning decisions are made, resulting in failure to entrench participatory democracy. The study found that Habermas (1979; 1987) validity claims have not been met because planning language is incomprehensible, insincere, untruthful and illegitimate, thereby compromising participation and inclusive urban planning in

Mzuzu City. First, this study revealed that the validity claim of comprehensibility was met at the block and ward level and community participation is high, because planners use the local language to communicate, but the validity claim of comprehensibility was not met in the service committees at city level and participation was compromised because planners use English, fraught with technical language and planning jargons, as the official language of communication. Second, the study revealed that the validity claims of sincerity was not met because the levels of planners' deception, which impede genuine flow of communication in service committees at city level, were high. While planners speak and write in a manner that appears to meet Habermas (1979; 1987) validity claims of sincerity, participants revealed episodes of deception and dishonesty during budgeting, resource allocation and the selection of project contractors in absence of community representatives. Third, the study unveiled that the validity claim of truthfulness was not met because the levels of cheating and factual inaccuracies in planners' oral and written communication, which impedes a genuine flow of planning ideas, knowledge and information, is high. While on the surface, planners' oral and written communication sounds as though they meet Habermas validity claim of truthfulness, but participants narrated episodes of lies, cheating and inaccuracies that arise in the service committees at City level, especially during budgeting, resource allocation, and determination of planning applications for development permission and in writing of the certificate of escalation.

Fourth, while both spoken and written language during planning meetings and in the UDP and USP documents meet Habermas (1979; 1987) validity claim of legitimacy, in that they comply with the legal provisions outlined in the Malawi Constitution (1995), the Local Government Act (1998 amended 2024) and the National Decentralization Policy (1998 amended 2024), however, participants complained that problems arise in the service committees and during the implementation phase. They revealed that what is legitimately communicated is not what usually gets implemented. It was also found that although key planning texts (UDP and USP) are sincerely, truthfully and legitimately written, but these documents have not been translated into the local language for everyone to read. Many communities are not aware of their existence.

Therefore, planning language fails to enhance community participation and inclusivity in the decision-making process in urban planning due to language barrier, due to lack of inclusiveness owing to the language barrier at city level where final planning decisions are made, resulting in failure to entrench participatory democracy

*ramework for Inclusive Participation in Urban Planning*

In order to realize the intentions of the policy and law in local governance, a framework that enables easy communication and understanding is proposed. Habermas (1979; 1987) requires the meeting of the validity claims of comprehensibility, sincerity, truthfulness and legitimacy. As communication is itself a precondition for democracy, poor communication implies lack of participation in the democratic decision-making process.

Therefore, to achieve full participation for effective inclusion in planning processes, certain conditions have to be met in Mzuzu City. These include**, the motivations and conditions for inclusive participation** for the purposes of realizing sustainable development goals (SDG11) as well as regional (Africa 2063) and national aspirations (mw2063). The study proposes a framework for inclusive community participation in planning. This framework has *four tiers* of community participation in urban planning. The first tier begins at the **Block Level**. This is where grassroots communities directly participate in the planning process. This is the first stage in the planning process. The **motives** of participation is that grassroots communities should identify and select community development projects and send them to the Neighbourhood Committees. The **condition** for inclusive participation should be that block leaders ensure that the identified projects truly and genuinely reflect the needs of the grassroots communities, rather than the ulterior motives of the community leaders. The **results** of participation of the grassroots communities must be projects that communities really need.

The second tier of participation is the **community planning committees**, split into two: Neighbourhood and Ward committees. The form of participation in this tier is indirect participation by elected members who participate on behalf of their people. The **motive** of participation must be to prepare area action plans which reflect the needs of communities at block levels. The **conditions** for inclusive community participation include: that whatever the members say and do should always reflect the true and genuine aspirations, needs and will of the grassroots communities. The **results** of participation, should be the action area plans which truly reflect the will of the grassroots communities, rather than the selfish needs of the community leaders and representatives.

The third level of participation is the **Council Service Committees**. This category should have both direct (planners/technocrats) and indirect (councillors representing communities and other stakeholders) participation. Participants must include; planners, councillors, other government officials; representatives of other interest groups. The **motives** of participation for planners should be to make planning, budgeting and funding decisions that advance the best interests of the grassroots communities, to provide technical, advice and orientation to councillors to enable them to ably represent their communities, to provide a conducive environment for councillors to fully participate in all decision-making processes of the service committees. The **motives** for the councillors should be to represent and amplify the voices of the grassroots communities, participate in the decision-making processes on the behalf of the communities, participate in budgeting and funding allocation and play a significant role in the selection of project contractors in the IPC. The **conditions** for effective participation should include: planners providing good technical advice and adequate orientation to councillors; planners sharing decision-making powers with community representatives; communities from all 15 wards must be represented by their councillors, not like it is right now where there only three councillors in service committees; and ensure that councillors fully participate in the selection process of the project contractors, to ensure transparency and accountability. The **results** of participation must indicate that the final decisions regarding plans and budgets must reflect the will of the communities, rather than the will of the technocrats and the councillors.